\documentstyle[aps,epsf,rotate,multicol]{revtex}
\begin{document}
\draft

\title{Impurity-induced diffusion bias in epitaxial growth}

\author{ Lu\'{\i}s A. Nunes Amaral$^{1}$ and Joachim Krug$^{2}$ }

\address{$^{1}$ Theorie II, Institut f\"ur Festk\"orperforschung, 
                Forschungszentrum J\"ulich, D-52425 J\"ulich, Germany \\ 
         $^{2}$ Fachbereich Physik, Universit\"at-GH Essen, 
                D-45117 Essen, Germany }

\date{September 9, 1996}

\maketitle

\begin{abstract}

  We introduce two models for the action of impurities in epitaxial
  growth.  In the first, the interaction between the diffusing adatoms
  and the impurities is ``barrier''-like and, in the second, it is
  ``trap''-like.  For the barrier model, we find a symmetry breaking
  effect that leads to an overall down-hill current.  As expected,
  such a current produces Edwards-Wilkinson scaling.  For the trap
  model, no symmetry breaking occurs and the scaling behavior appears
  to be of the conserved-KPZ type.

\end{abstract}


\begin{multicols}{2}

The effect of impurities on growth rate and morphology is a classic
topic of crystal growth theory \cite{cabrera}. The most thoroughly
studied case is the step flow growth of a vicinal surface, when the
(immobile) impurities pin the advancing steps and thus lead to step
bunching \cite{hmk,kandel}. These theories are {\em mesoscopic} rather
than microscopic in nature, in the sense that they describe the
interaction of preexisting steps with discrete impurities
\cite{kandel} or an impurity concentration field \cite{hmk}.

The advent of modern crystal growth techniques aimed at manufacturing
layers of atomic scale thickness, notably molecular beam epitaxy (MBE)
\cite{mbe}, has lead to a renewed appreciation of the fact that small
concentrations of impurity atoms on the growing surface can
drastically influence the growth kinetics. A particularly striking
aspect of these recent results is that the impurities may either lead
to a deterioration of the growth quality -- as would be expected
according to the classic view \cite{cabrera} -- or, conversely, they
may play the role of {\em surfactants} in stabilizing smooth,
layer-by-layer growth \cite{voigtlaender}.

A clear example of the former type is the effect of hydrogen on the
MBE of silicon \cite{eaglesham,adams}.  It was observed that the
presence of H in the growth chamber during Si deposition on Si(001)
leads to a decrease of the epitaxial height \cite{eaglesham}, at which
epitaxy breaks down and the growth becomes amorphous, proportional to
the logarithm of the partial pressure of H \cite{adams}. The
experiments ruled out the hypothesis that the breakdown of epitaxy
might be due either to an increased coverage of H at the interface
\cite{eaglesham} or to its incorporation into the bulk \cite{copel}.
Rather, it was concluded \cite{adams} that the hydrogen greatly speeds
up the development of surface roughness due to a reduction of the
diffusion length of Si adatoms \cite{vasek}.

The modification of the diffusion properties of the adatoms appears to
be the most significant effect of the impurities also when they act as
surfactants \cite{zhang,kaxiras}, though the nature of the
modification --- for example, whether the diffusion length is
increased or decreased --- depends on the chemical species in a
complicated way \cite{voigtlaender}.  This sensitivity to atomic
details is rather unexpected, and calls for the development of models
which are more microscopic than previous approaches \cite{hmk,kandel}.

In this Letter, we introduce two models for the action of impurities
in epitaxial growth. Rather than attempting a detailed description
of some particular material, our aim is to define a `minimal' model in
which the consequences of the impurity-adatom interaction on the large
scale morphology of the surface can be clearly elucidated.  The study
of oversimplified models \cite{wolf,dassarma} of `ideal MBE'
\cite{lai} has previously been very successful in clarifying the
universality classes for kinetic roughening \cite{lai,villain,tang}
and morphological instability \cite{villain,siegert,kps} in the
absence of impurities \cite{reviews}.

Our models reproduce the sensitive dependence on microscopic details
mentioned above: Using two equally plausible microscopic interaction
mechanisms -- of {\em barrier}-type and of {\em trap}-type,
respectively -- we find that for the barrier model the impurities
neutralize the destabilizing effect of step edge barriers
\cite{villain,kps} and thus lead to smoother growth \cite{markov},
while for the trap model the asymptotic morphology remains
unaffected. In terms of the coarse-grained continuum description of
the surface \cite{villain,reviews} we are able to trace the difference
between the two models to the fact that the barriers modify the {\em
symmetry} of the surface diffusion process, while the traps do not.

\begin{figure}
\narrowtext
\centerline{
  ~~\epsfysize=.9\columnwidth{{\epsfbox{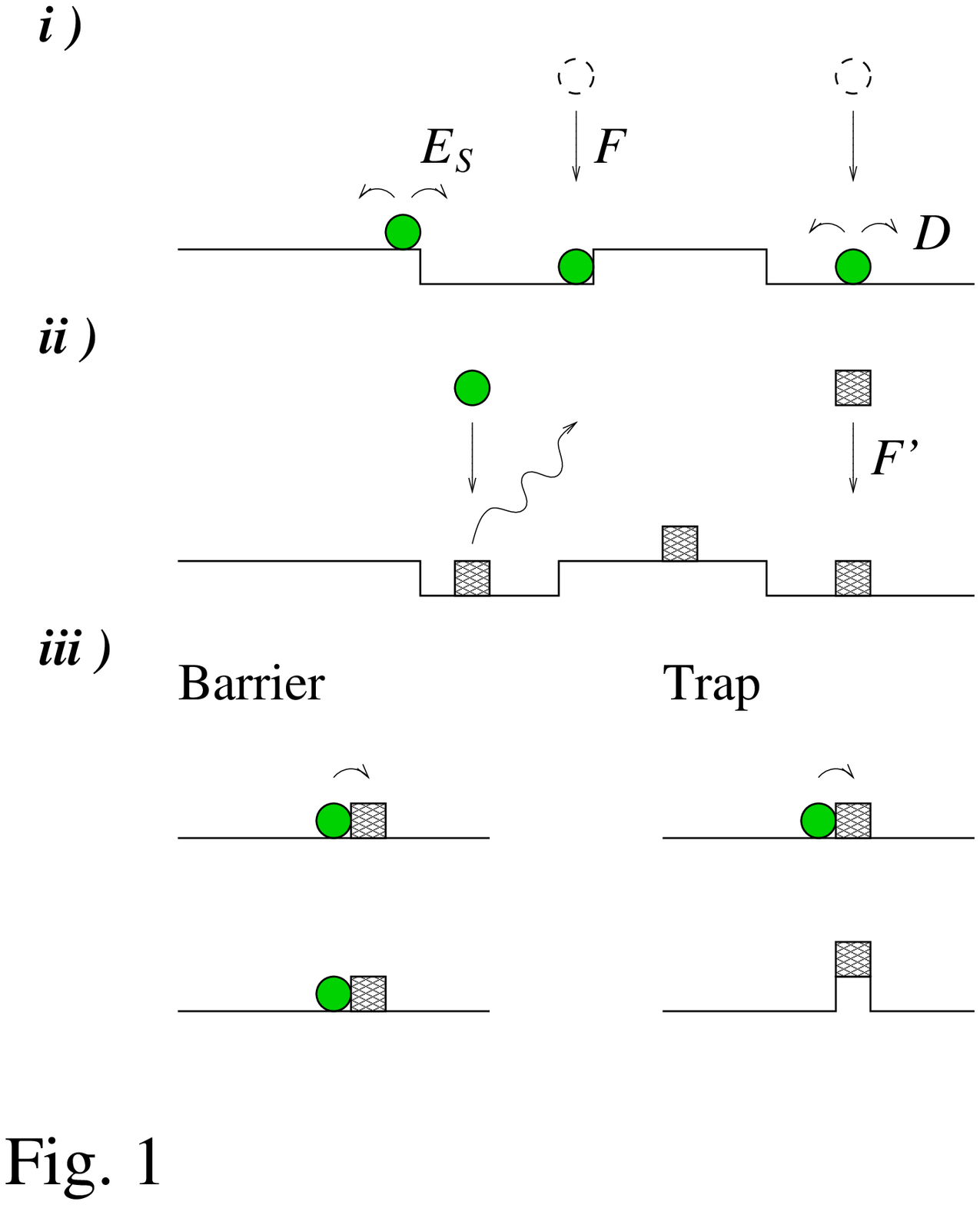}}}
}
\vspace*{0.5cm}
\caption{
  Schematic representation of the processes present in our models.
  Material is deposited at a rate $F$ and single atoms
  (gray circles) diffuse at a rate $D$.  At step edges,
  adatom motion down the step is accepted with probability $p =
  \exp(-E_S)$, where $E_S$ is the step edge barrier.  Impurities
  (squares) are deposited at a rate $F'$ and are not allowed to
  diffuse.  When an atom is deposited over an impurity, the latter
  evaporates and is replaced by the deposited atom.  Impurities can
  interact with the adatoms through two distinct mechanism.  
  In the barrier-like interaction, an adatom
  trying to diffuse on top of an impurity will have its move rejected.
  In the trap-like interaction, the same adatom would have been
  trapped, i.e., it would swap positions with the impurity and stop
  diffusing.
}
\label{f-models}
\end{figure}

The models proposed in this Letter have three main ingredients (see
Fig. \ref{f-models}).

{\it i) Deposition and Diffusion:} For simplicity, we consider a
one-dimensional discrete substrate.  Material is randomly deposited at
a rate $F$. The deposition occurs in a solid-on-solid (SOS) manner,
i.e. deposition at a position $i$ implies that the surface height
$h(i)$ is increased by one unit. Every atom which has only one
occupied neighbor (namely, in the layer below) is considered a mobile
{\it adatom\/} which diffuses at a rate $D$; atoms with more bonds are
immobile.  When an adatom diffuses to a step edge from above we
implement an additional energy barrier \cite{schwoebel} by accepting a
diffusion move down the step only with probability $p = \exp (- E_S
)$, where the barrier energy $E_S$ is measured in units of $k_B T$. In
the absence of impurities such step edge barriers are know to lead to
unstable growth \cite{villain,kps,reviews}.

{\it ii) Impurities:} Based on the experimental observations described
previously, we assume that there is a flux $F'$ of impurities onto the
growing surface.  We restrict our study to the limit in which the
diffusion rate of the impurities is much smaller than the diffusion
rate of the adatoms, so that the impurities can be considered
immobile. We also assume that impurities evaporate from the surface at
a rate that keeps $\theta_I$, the impurity coverage, approximately
constant.  This is done by removing the impurity whenever a new atom
is deposited on top of it.  An important consequence of this rule is
that the average lifetime of an impurity at a given site equals the
monolayer deposition time.

{\it iii) Interactions:} Concerning the interactions between the
impurities and the diffusing adatoms, we introduce two alternative
models: the barrier and the trap model.  In the {\it barrier\/} model,
an adatom trying to diffuse onto a site occupied by an impurity will
have its attempted move rejected.  In the {\it trap\/} model, the same
adatom will move on top of the impurity and then will swap positions.
The end result is that it will no longer be able to diffuse because it
has (at least) two ``chemical bonds'': to the atom below and to the
impurity.

\begin{figure}
\narrowtext
\centerline{
  ~~\epsfysize=.9\columnwidth{\rotate[r]{\epsfbox{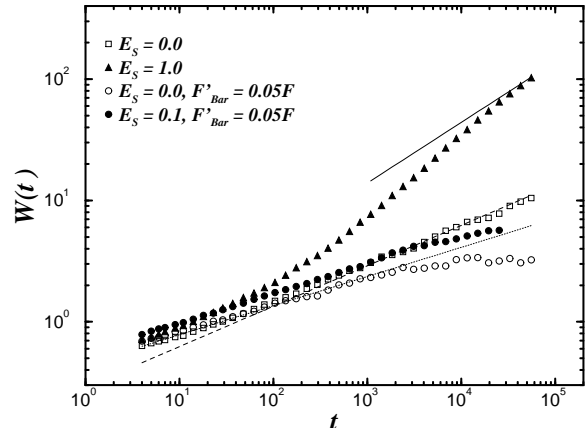}}}
}
\vspace*{0.5cm}
\caption{
  Plot of the total surface width as a function of time
  (or coverage) for barrier-like impurities.
  The results shown were obtained for $L=1000$ and $D/F = 10^4$.
  Averages were taken over 50 runs.  In the absence of step edge
  barriers or impurities $W(t)$ diverges with an exponent $\beta \approx
  1/3$.  When step edge barriers are introduced, the instability in the
  growth process leads to an effective exponent close to one at intermediate
  times and the random deposition value $\beta \approx 0.5$ at long
  times.  The presence of even a small amount of impurities leads to a 
  significantly smoother surface. The lines are plotted as guides to
  the eye, and have slopes $0.24$ (dotted), $0.33$ (dashed), and
  $0.50$ (full).
}
\label{f-barrier}
\end{figure}

In our simulations we focus on the exponent $\beta$ describing the
increase of the surface width \cite{reviews} $W(t,L) \equiv \langle (h
- \langle h \rangle)^2\rangle^{1/2} \sim t^\beta$ in the early time
regime $t \ll L^z$; here $L$ denotes the system size and the dynamic
exponent is $z = 1/(1-2\beta)$ for the class of (one-dimensional)
models considered in this paper \cite{wolf}. To put the results into
perspective, we will compare them to the predictions of the
appropriate continuum equations for the coarse-grained height function
$h(x,t)$; as usual, the average height $\langle h \rangle = Ft$ will
be subtracted. In the present context the following equation suffices
\cite{lai,villain,tang}:
\begin{equation}
\label{geneq}
\frac{\partial h}{\partial t} = \nu \nabla^2 h - \frac{\lambda}{2}
\nabla^2 (\nabla h)^2 - \kappa (\nabla^2)^2 h + \eta,
\end{equation}
where the stochastic force $\eta(x,t)$ models the shot noise in the 
beam, and can be taken to be Gaussian with zero mean and covariance
\begin{equation}
\label{covariance} 
\langle \eta(x,t) \eta(x',t') \rangle =  F \delta(x - x') \delta(t-t').
\end{equation}
The first term on the right hand side of (\ref{geneq}) arises from the
gradient expansion of an inclination-dependent, growth induced surface
current \cite{villain,kps}.  When it is present, it dominates the
large scale morphology: For $\nu > 0$ one obtains kinetic roughening
of the Edwards-Wilkinson (EW) universality class \cite{reviews,EW}
with $\beta = 1/4$, while for $\nu < 0$ the growth is unstable and a
mound morphology is expected to develop \cite{siegert,reviews,mounds}.
If, for reasons of symmetry (see below) $\nu = 0$, the second,
nonlinear term becomes important, and changes the roughening exponent
to $\beta = 1/3$ \cite{villain,lai,tang} (the ``conserved
Kardar-Parisi-Zhang'' universality class \cite{KPZ}).  Finally, in
many cases the growth-induced coefficients $\nu$ and $\lambda$ are
small, and the early time behavior is dominated by the third term in
(\ref{geneq}), which arises from equilibrium surface diffusion
\cite{mullins} and leads to a (transient) value $\beta = 3/8$
\cite{wolf,dassarma}.

We consider first the case of a nonzero step edge barrier in the
absence of impurities; then $\nu < 0$ \cite{villain,kps} and one
expects asymptotically unstable growth.  As shown in Fig.
\ref{f-barrier}, the instability sets in after an initial power law
transient which terminates at about 100 monolayers.  Later on, the
destabilizing effect of the step edge barrier leads to wavelength
selection and mound formation with a very rapid growth of the surface
width.  Once large slopes have appeared on the surface, there is
hardly any transfer of matter between the different mounds and the
exponent $\beta$ reaches the limiting value $\beta = 1/2$
characteristic of random deposition \cite{lanczycki}.

\begin{figure}
\narrowtext
\centerline{
  ~~\epsfysize=.9\columnwidth{\rotate[r]{\epsfbox{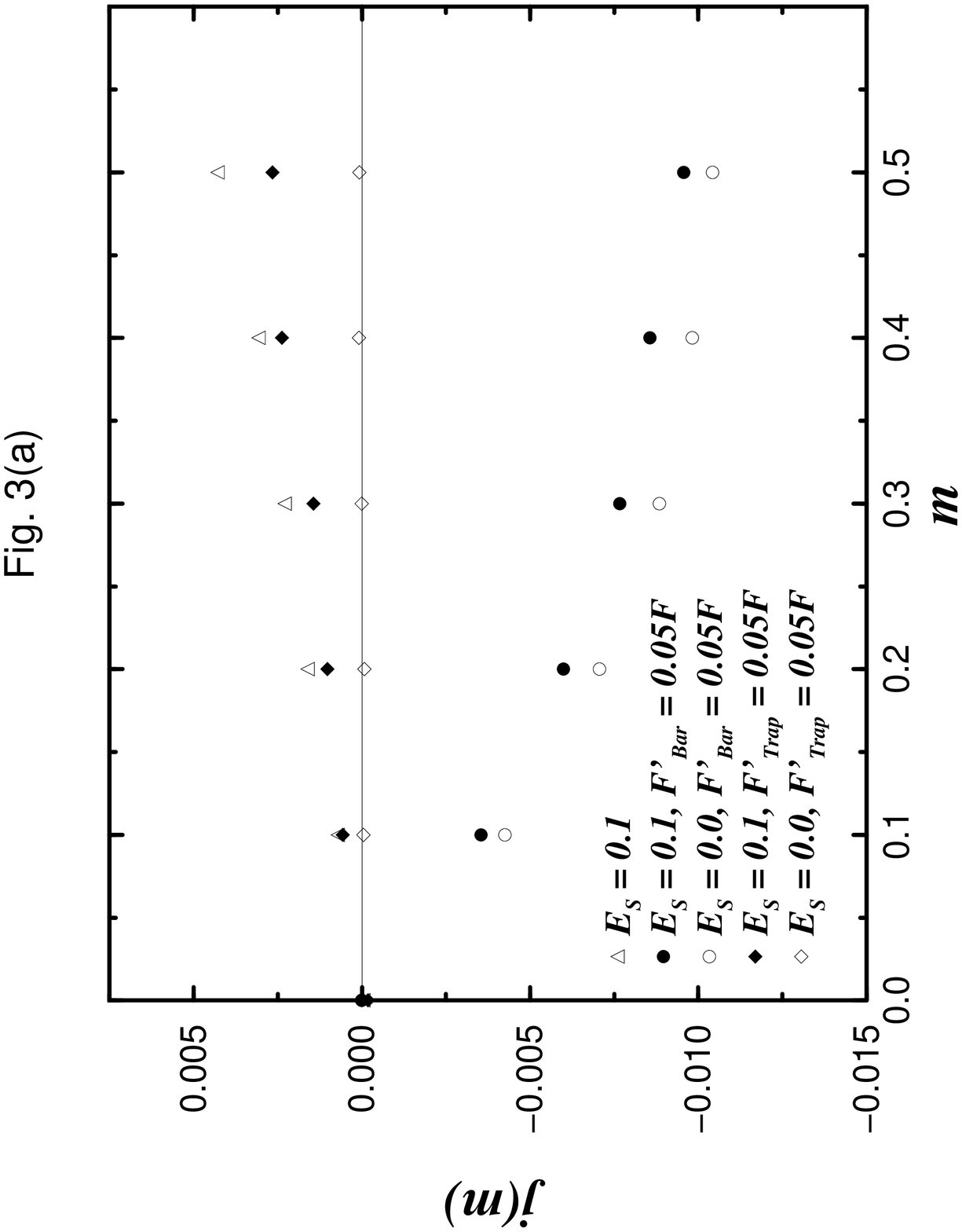}}}
}
\vspace*{0.7cm}
\centerline{
  ~~\epsfysize=.9\columnwidth{{\epsfbox{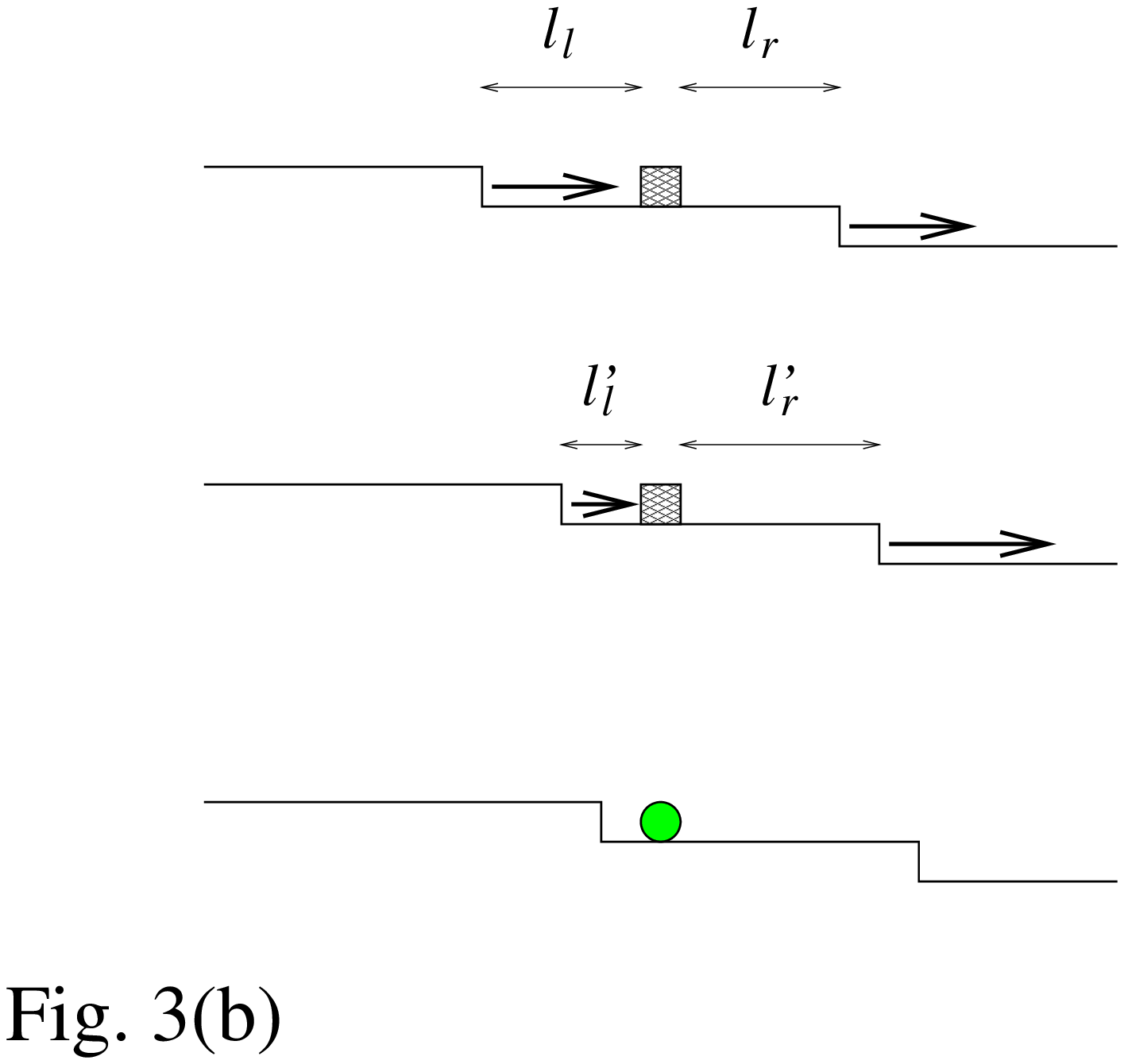}}}
}
\vspace*{0.5cm}
\caption{
  a) Plot of the current $j(m)$ as a function of the tilt $m$ of the
  interface.  The tilt is prescribed numerically through helicoidal
  boundary conditions.  Our results make it clear that the
  barrier-like impurities lead to a negative current even in the
  presence of a step edge barrier, while the trap impurities leave the
  sign of the current unaffected.  We used a smaller value of the step
  edge barrier because of the numerical difficulties in calculating
  the current (see Ref.~ \protect\cite{kps}). b) Schematic
  representation of the effect of a barrier-like impurity on the
  motion of a step train.  On average an impurity will be deposited on
  the middle of a terrace.  However, while the upper step moves
  towards the impurity, the lower step moves away from it, so that in
  fact $l_l < l_r$, where $l_l$ ($l_r$) is the length of the portion
  of the step to the left (right) of the impurity. Since the up-hill
  current is proportional to $l_l$ and the down-hill current is
  proportional to $l_r$, we will have an average down-hill current
  which leads to a positive $\nu$ coefficient and a stable interface.
  Note that this mechanism can only be effective if during the time to
  deposit one monolayer the impurity will have been removed from its
  position; otherwise, the impurity would pin the step and destabilize
  the surface.
}
\label{f-current}
\end{figure}

When we introduce impurities of the barrier-type a striking change
occurs.  As is visually apparent from Fig. \ref{f-barrier}, the
presence even of small amounts of barrier impurities leads to a
significative decrease of the interface width. The exponent takes the
value $\beta = 0.24 \pm 0.03$, consistent with EW universality.  The
natural interpretation is that the impurities have caused the
coefficient $\nu$ in Eq.~(\ref{geneq}) to change sign, from
destabilizing ($\nu < 0$) to stabilizing ($\nu > 0$). The value of
$\nu$ can be directly ascertained by measuring the average surface
current for tilted substrates \cite{kps}. The results, shown in Figure
\ref{f-current}(a), confirm our interpretation: Even a small flux of
impurities ($F'/F = 0.05$) leads to a sizable positive value of $\nu$,
both in the presence of a step edge barrier and for $E_S = 0$.

The reason for the change produced by the barrier impurities can be
understood as follows.  As is well known \cite{villain,kps}, the step
edge barrier leads to an up-hill current because adatoms are rejected
when trying to go down step edges and become integrated in the bulk
when reaching an ascending step.  This difference leads to an average
current towards the up-step which destabilizes the surface, as
described earlier.  To visualize the effect of the impurities on this
process, let us consider a step train moving from left to right; cf.
Fig \ref{f-current}(b).  An impurity can be deposited anywhere on a
given terrace, so we can say that on average it is deposited in the
{\it middle\/} between the two steps.  However, as more material is
deposited, the step edge to the left of the impurity advances towards
it.  On the other hand, the step edge to the right of the impurity
moves away from it.  Thus, the distance to the step to the left of the
impurity is typically {\em smaller} than the distance to the right.
Since the current away from the impurity on each side is proportional
to the material deposited there (and thus to the length of that part
of the terrace) we see that an average {\it down-hill\/} current is
generated.

\begin{figure}
\narrowtext
\centerline{
  ~~\epsfysize=.9\columnwidth{\rotate[r]{\epsfbox{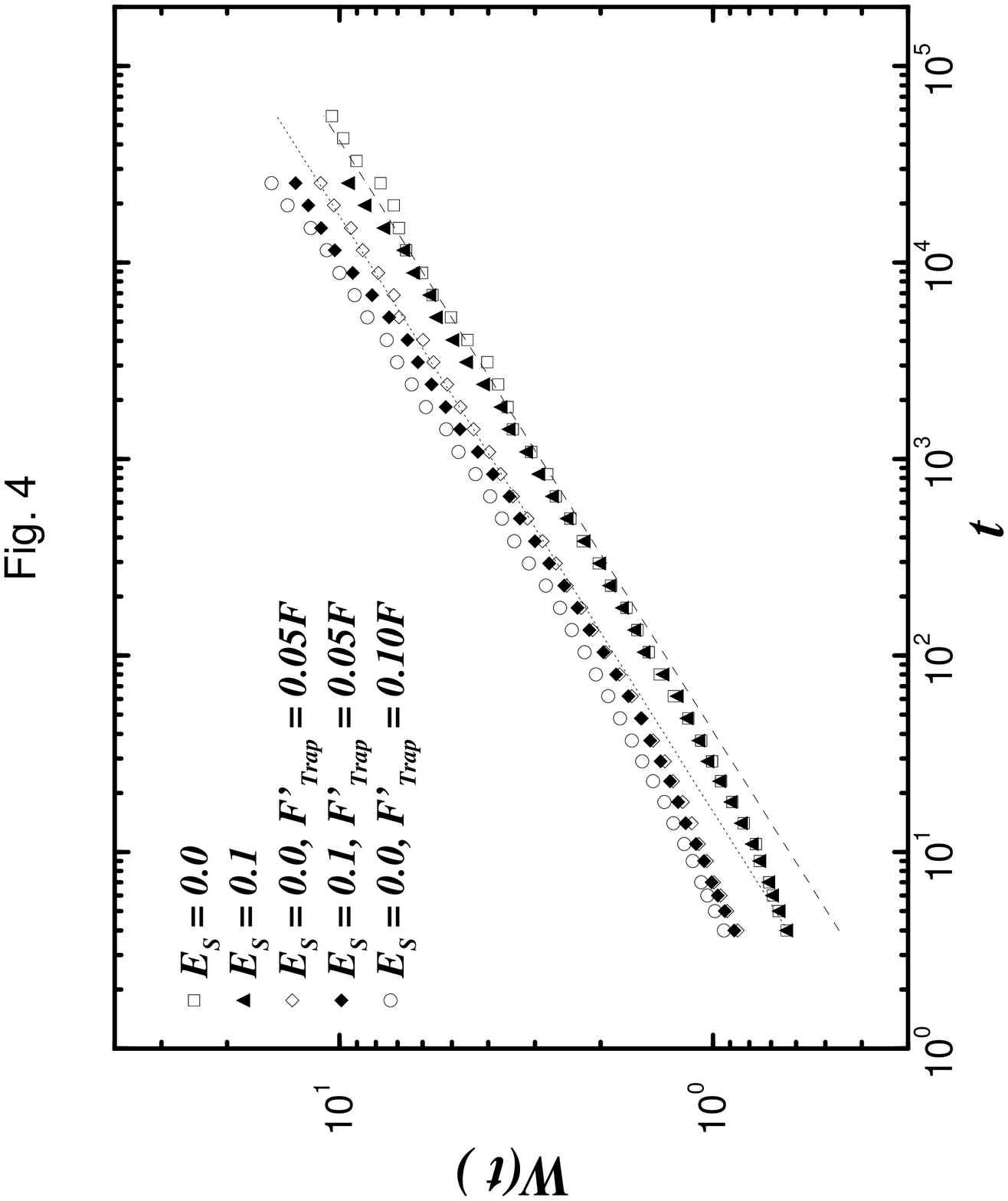}}}
}
\vspace*{0.5cm}
\caption{
  Plot of the total surface width as a function of time for trap-like
  impurities, under the same conditions as in
  Fig.~\protect\ref{f-barrier}. In this case, the effect of the
  impurities is not as striking as for the barrier-like impurities.
  Even in the presence of step edge barriers, the main effect seems to
  be an increase in the prefactor of the width.  The straight lines
  are plotted as guides to the eye, and have slope $0.33$.
}
\label{f-trap}
\end{figure}

In contrast, the trap impurities do not seem to significantly change
the dynamics of the growth process, apart from an increase of the
prefactor of the width (Fig.~\ref{f-trap}); certainly they are not
able to suppress the destabilizing effect of the step edge barriers.
This is confirmed by a measurement of the surface current, which
remains up-hill in the presence of traps, as shown in Fig.
\ref{f-current}(a).  

The traps do not induce a surface current because they cannot {\em
bias} the diffusion of adatoms: The trap impurity makes itself felt
only when the atom has already jumped onto it, and therefore no longer
participates in the mass transport on the surface (when the impurity
disappears due to the deposition of an additional atom, the trapped
adatom remains immobile).

This point is brought out more clearly by considering the trap model
without step edge barriers. For $E_S = 0$, the pure model ($F'=0$) has
a symmetry which forces $\nu = 0$ in Eq.~(\ref{geneq}): For any local
environment, the probability of a mobile adatom to jump to the right
is equal to that for a jump to the left. Since this is true
irrespective of the overall surface tilt, no growth-induced current
can exist \cite{symmetry}. With $\nu = 0$ the behavior of
(\ref{geneq}) is dominated by the second, nonlinear term, and one
expects $\beta = 1/3$. Our simulations lead to a value of $\beta =
0.33 \pm 0.03$, indicating that this symmetry is preserved by the trap
impurities.

The increase in the prefactor of the width with increasing trap
concentration can be interpreted as a decrease of the diffusion
length $\ell_D$: Indeed, it can be shown \cite{brendel96} that
the prefactor scales as $\ell_D^{-4d/(10+d)}$ for a $d$-dimensional
surface. {}From the data shown in Fig.~\ref{f-trap} we therefore
estimate that an impurity flux $F'/F = 0.1$ decreases the diffusion
length by almost a factor of three.

Finally, one may ask why none of our models shows an {\em
  impurity-induced} growth instability of the kind considered in the
classic theories of step bunching \cite{hmk,kandel}. We believe that
this is due to the fact that the lifetime of an impurity at a given
position is, in our models, fixed to be of the order of the monolayer
deposition time. In terms of the conventional step flow picture
\cite{cabrera,hmk} it is evident that impurities can effectively pin
steps only if they remain at a position much longer than the time
required for a step to pass over a terrace. Thus, it appears important
to consider models with a variable lifetime for the impurities.

We acknowledge stimulating discussions with M. Rost, M. Schimschak and
P. \v{S}milauer.

\end{multicols}


\begin{thebibliography}{10}

\bibitem{cabrera} 
N. Cabrera and D.~A. Vermilyea, in {\em Growth and Perfection of 
Crystals}, ed. by R. Doremus, B. Roberts and D. Turnbull
(Wiley, New York 1958), p. 393.

\bibitem{hmk}
J.~P. v.d. Eerden and H. M\"uller-Krumbhaar, Phys. Rev. Lett. {\bf 57},
2431 (1986).

\bibitem{kandel}
D. Kandel and J.~D. Weeks, Phys. Rev. B {\bf 49}, 5554 (1994).

\bibitem{mbe}
{\it Molecular Beam Epitaxy}, ed. by A. Cho, (AIP Press, Woodbury,
NY 1994).

\bibitem{voigtlaender} 
B. Voigtl\"ander, A. Zinner, T. Weber and H.~P. Bonzel, Phys. Rev. B
{\bf 51}, 7583 (1995), and references therein.

\bibitem{eaglesham} 
D.~J. Eaglesham, J. Appl. Phys. {\bf 77}, 3597 (1995).

\bibitem{adams}
D.~P. Adams, S.~M. Yalisove, and D.~J. Eaglesham, Appl. Phys. Lett.
{\bf 63}, 3571 (1993).

\bibitem{copel}
M. Copel and R.~M. Tromp, Phys. Rev. Lett. {\bf 72}, 1236 (1994).

\bibitem{vasek}
J.~E. Vasek, Z. Zhang, C.~T. Salling, and M.~G. Lagally, Phys. Rev.
B {\bf 51}, 17207 (1995).

\bibitem{zhang} 
Z. Zhang and M.~G. Lagally, Phys. Rev. Lett. {\bf 72}, 693 (1994).

\bibitem{kaxiras}
D. Kandel and E. Kaxiras, Phys. Rev. Lett. {\bf 75}, 2742 (1995).

\bibitem{wolf}
D.~E. Wolf and J. Villain, Europhys. Lett. {\bf 13}, 389 (1990).

\bibitem{dassarma}
S. Das Sarma and P. Tamborenea, Phys. Rev. Lett.  {\bf 66}, 325
(1991); P.~I. Tamborenea and S. Das Sarma, Phys. Rev. E {\bf 48},
2575 (1993).

\bibitem{lai}
Z.-W. Lai and S. Das Sarma, Phys. Rev. Lett. {\bf 66}, 2348 (1991).

\bibitem{villain} 
J. Villain, J. Phys. France I {\bf 1}, 19 (1991).

\bibitem{tang}
L.-H. Tang and T. Nattermann, Phys. Rev. Lett. {\bf 66}, 2899 (1991).

\bibitem{siegert}
M. Siegert and M. Plischke, Phys. Rev. Lett. {\bf 68}, 2035 (1992);
Phys. Rev. E {\bf 50}, 917 (1994).

\bibitem{kps}
J. Krug, M. Siegert, and M. Plischke, Phys. Rev. Lett. {\bf 70}, 3271
(1993).

\bibitem{reviews}
For reviews see A.-L.  Barab\'asi and H.~E.  Stanley, {\it Fractal
Concepts in Surface Growth\/} (Cambridge University Press, Cambridge,
1995); J. Krug, Adv. Phys. (in press).

\bibitem{markov}
A related mechanism has been described by I. Markov, Phys. Rev. B 
{\bf 50}, 11271 (1994).

\bibitem{schwoebel}
G. Ehrlich and F.~G. Hudda, J. Chem. Phys. {\bf 44}, 1039 (1966);
R.~L. Schwoebel and E.~J. Shipsey, J. Appl. Phys. {\bf 37}, 3682
(1966).

\bibitem{EW}
S.~F. Edwards and D.~R. Wilkinson, Proc. R. Soc. Lond. {\bf A381}, 
17 (1982).

\bibitem{mounds}
P. \v{S}milauer and D.~D. Vvedensky, Phys. Rev. B {\bf 52}, 
14263 (1995).

\bibitem{KPZ}
M. Kardar, G. Parisi, and Y.-C. Zhang, Phys. Rev. Lett. {\bf 56}, 
889 (1986).

\bibitem{mullins}
W.~W. Mullins, J. Appl. Phys. {\bf 28}, 333 (1957).

\bibitem{lanczycki}
C.~J. Lanczycki and S. Das Sarma, Phys. Rev.  Lett. {\bf 76}, 780
(1996).

\bibitem{symmetry} 
For related symmetries in other models see J.Krug in
Ref.\cite{reviews}.

\bibitem{brendel96}
L. Brendel, H. Kallabis, J. Krug, M. Schroeder, J. Villain and
D.E. Wolf (unpublished); see also J. Krug in Ref.\cite{reviews}. 

\end{thebibliography}
\end{document}